\documentclass[aps,prc,twocolumn,superscriptaddress,nofootinbib]{revtex4-1}
\usepackage[T1]{fontenc}
\usepackage[latin9]{inputenc}
\usepackage{amstext}
\usepackage{graphicx}
\begin{document}

\preprint{This line only printed with preprint option}

\title{Ground state octupole correlation energy with effective forces}

\author{L.M. Robledo}

\email{luis.robledo@uam.es}

\homepage{http://gamma.ft.uam.es/robledo}

\affiliation{Departamento de F\'\i sica Te\'orica, Universidad Aut\'onoma de Madrid, E-28049
Madrid, Spain}

\begin{abstract}
The ground state octupole correlation energy is computed with the D1M 
variant of the Gogny force in different theoretical frameworks and 
analyzed in detail. First I consider the correlation energy gained at 
the mean field level by breaking reflection symmetry. Next I consider 
the energy gain coming from symmetry (parity) restoration and finally 
I analyze the ground state correlation energy after configuration 
mixing with axially symmetric octupole states. The impact of the latter 
on theoretical binding energies indicates that octupole correlations do 
not affect in a significant way the trend and systematic of binding 
energies and therefore can not improve the performance of  
theoretical models in this respect. In particular, the too-large ``shell gaps" 
predicted by self-consistent mean field models and relevant in 
astrophysics scenarios are not altered by the octupole correlations.
\end{abstract}
\maketitle

\section{Introduction}

Many physical applications require a precise and accurate knowledge of 
nuclear binding energies. A typical example is found in the area of 
astrophysical simulations used to describe explosive scenarios where 
the goal is to understand how chemical elements heavier than iron are 
formed in the Universe \cite{Arn.07,Arc.12,Rei.14}. The nuclear 
reactions involved take place along the "r-process" path that is 
located in regions of neutron rich nuclei well away from the stability 
line. Those nuclei are not within experimental reach in the present day 
or near future experimental facilities \cite{Bla.06} and therefore 
astrophysical simulations require as input theoretical nuclear 
structure data. A consistent framework working with the same accuracy 
all over the nuclide chart, as required by astrophysical simulations, 
necessarily involves mean field techniques with effective interactions 
\cite{Ben.03}. For the latter it is customary to used the family of 
Gogny forces \cite{Decharge.80} focusing on the D1M parametrization 
\cite{D1M} for its good reproduction of nuclear binding energies. All 
the Gogny forces are characterized by a finite range central potential 
linear combination of  gaussian shapes with different ranges and the 
standard combination of spin, isospin and spatial exchange terms. A 
standard zero range spin-orbit term and the Coulomb potential 
constitute the rest of the nuclear potential. The force is supplemented 
by a zero range density dependent term used to enforce the saturation 
mechanism of atomic nuclei. The parameters of the different Gogny 
forces are fitted using different strategies. In the case of D1M the 
parameters were fitted  to reproduce the binding energies of all known 
nuclei computed with the Hartree- Fock- Bogoliubov (HFB) method 
\cite{RS.80} supplemented with an approximate rotational energy 
correction and an approximate zero point energy correction from 
quadrupole motion in the spirit of the 5D Bohr hamiltonian. In this way 
an impressive rms deviation for the binding energy of 0.79 MeV is 
obtained. Charge radii and realistic symmetric and neutron matter 
equation of state were also used as  target quantities in the fit. The 
specific values of the D1M parameters are given in Ref \cite{D1M}. 
Recently, additional correlation energies including Projection on 
Particle Number (PNP), exact angular momentum projection (AMP) of 
deformed axially symmetric states and quadrupole configuration mixing 
with the projected states have been computed with Gogny D1M 
\cite{Tom.14}. Unfortunately none of these effects seem to improve the 
agreement with experimental data in a significant way at variance with 
other studies using Skyrme or Gogny forces \cite{sa07,de10}. 

Next in the hierarchy of multipole moments describing the shape of the 
nucleus comes the octupole moment with its characteristic breaking of 
reflection symmetry (parity). The purpose of this paper is to explore 
the correlation energy associated with the axially symmetric octupole 
degree of freedom \cite{bu96,Gaf.13} and its impact on binding 
energies. Contrary to the quadrupole case, where correlation energies 
reach tens of MeV already at the mean field level, the octupole ones 
never exceed 2.5 MeV. This is a modest addition but can have a strong 
impact on astrophysical scenarios where an accuracy of a couple hundred 
keV is required \cite{Arc.12}.

This paper is a follow up of Ref \cite{Rob.11b} where a systematic 
calculation of octupole properties over a large set of  even-even 
nuclei was carried out. In the accompanying material of \cite{Rob.11b} 
three tables with the relevant quantities obtained with three 
parametrizations of the Gogny force were included. In those tables the 
correlation energies were already included but never discussed. Given 
the renewed interest on nuclear binding energies 
\cite{Arc.12,Rei.14,Tom.14} I consider now timely to discuss in detail 
the impact of octupole correlations in those quantities. 

\section{Methodology}

\begin{figure}[htb]
\includegraphics[width=8.5cm]{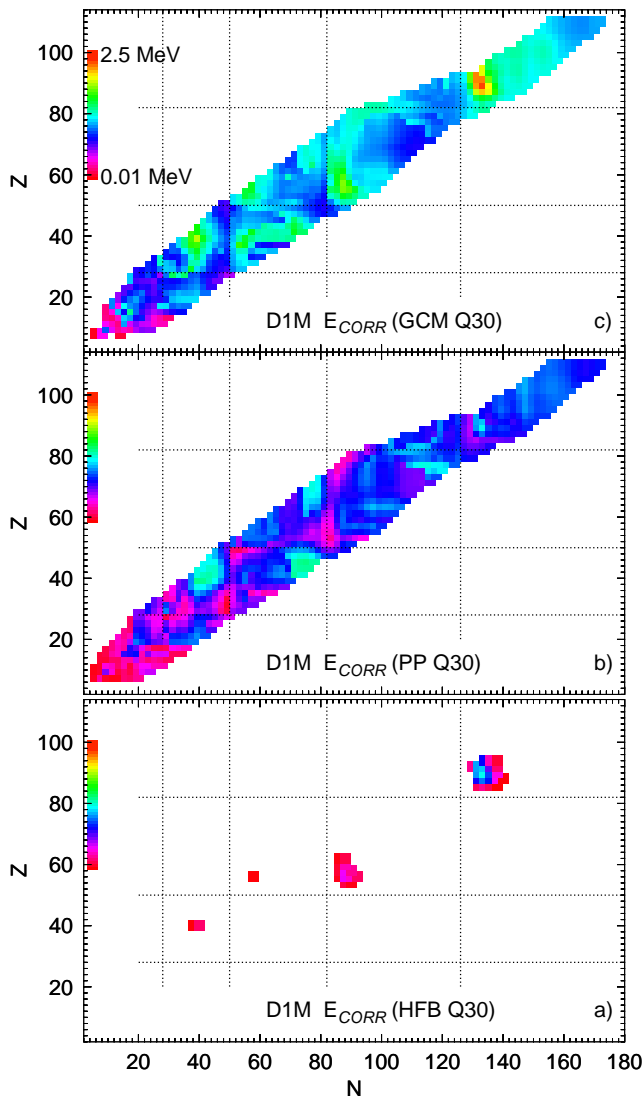}
\caption{(Color online) Color scale representation of the octupole correlation energy gain
as compared to  HFB results preserving reflection symmetry.
In panel a) the HFB correlation energy gained by breaking reflection symmetry.
In panel b) the parity RVAP correlation energy. In panel c) the octupole
GCM correlation energy. Horizontal and vertical dotted lines correspond
to magic proton and neutron numbers.\label{fig-1}}
\end{figure}

The computational procedure is the same as in Ref \cite{Rob.11b} where 
a thorough account of the properties of negative parity excited states 
was given using the HFB, parity projection and configuration mixing 
methods.  Therefore I will give here only a short description for the 
convenience of the reader. First, the HFB equation with a constraint on 
the axial octupole moment is solved using the approximate second order 
gradient method \cite{Rob.11a} with the HFBaxial computer code 
\cite{HFBaxial}. A set of axially symmetric HFB wave functions $|\Phi 
(Q_{30})\rangle$ is generated this way. The interaction used is the 
Gogny force \cite{Decharge.80} with the D1M parametrization \cite{D1M}. 
The location of the minimum of the HFB energy curve $E_\textrm{HFB} 
(Q_{30})$ determines whether the ground state is octupole deformed or 
not. The energy difference $\Delta E_\textrm{HFB} = E_\textrm{HFB} 
(Q_{30\, \textrm{min}})-E_\textrm{HFB} (0)$ is the mean field 
correlation energy gained by the spontaneous breaking of reflection 
symmetry. In the next step, the octupole deformed HFB states are 
projected to good parity $\pi = \pm 1$ and the corresponding projected 
energies 
\begin{equation}
E_\pi (Q_{30}) = \frac{\langle \Phi(Q_{30}) |\hat{H} \hat{P}_\pi | \Phi(Q_{30})\rangle }{\langle \Phi(Q_{30})|\hat{P}_\pi |  \Phi(Q_{30})\rangle }
\end{equation}
are computed (paying special attention to the consistency problems 
associated with the use of density dependent forces 
\cite{Rob.07,Rob.10}). The minima of the two curves $E_+(Q_{30})$ and 
$E_-(Q_{30})$ at octupole moment values $Q_{30, \textrm{PP min}}$ and 
$Q_{30, \textrm{NP min}}$, respectively, determine the optimal 
intrinsic state for each parity in the spirit of the restricted 
variation after projection (RVAP) method \cite{Egi.91}. In even-even 
nuclei the ground state always has positive parity and the RVAP 
correlation energy is given by $\Delta E_\textrm{RVAP} = E_+ (Q_{30, 
\textrm{PP min}})-E_\textrm{HFB} (0)$. Finally, a Generator Coordinate 
Method (GCM) calculation is performed using the set  $|\Phi 
(Q_{30})\rangle$ of HFB wave functions as basis (generating coordinate) 
states. The solution of the Hill-Wheeler equation corresponding to the 
minimum energy  is by definition the ground state, that has positive 
parity in even-even nuclei. The correlation energy is defined in this 
case as $\Delta E_\textrm{GCM} = E_\textrm{GS GCM}-E_\textrm{HFB} (0)$. 

By considering the octupole degree of freedom alone, I am implicitly 
assuming  that it is not coupled to other degrees of freedom like 
quadrupole deformation or pairing. Given the negative parity of this 
degree of freedom, the above assumption seems reasonable but it has yet 
to be confirmed. In similar GCM calculations coupling the quadrupole 
and octupole degree of freedom \cite{Rob.88,Mey.95,Rod.12,Rob.13} this 
assumption seems to be fulfilled and, as a consequence, the ground 
state octupole correlation energy is an additive quantity.

Recently, non axial degrees of freedom $Q_{32}$ related to tetrahedral 
correlations have been considered by several authors both with Skyrme 
\cite{Zbe.06} and Gogny \cite{Tag.15} forces. Their impact on ground 
state correlation energies over the whole nuclear chart and specially 
when angular momentum projection is considered \cite{Tag.15} remains to 
be explored.

\section{Results}

\begin{figure*}
\includegraphics[width=10.5cm,angle=270]{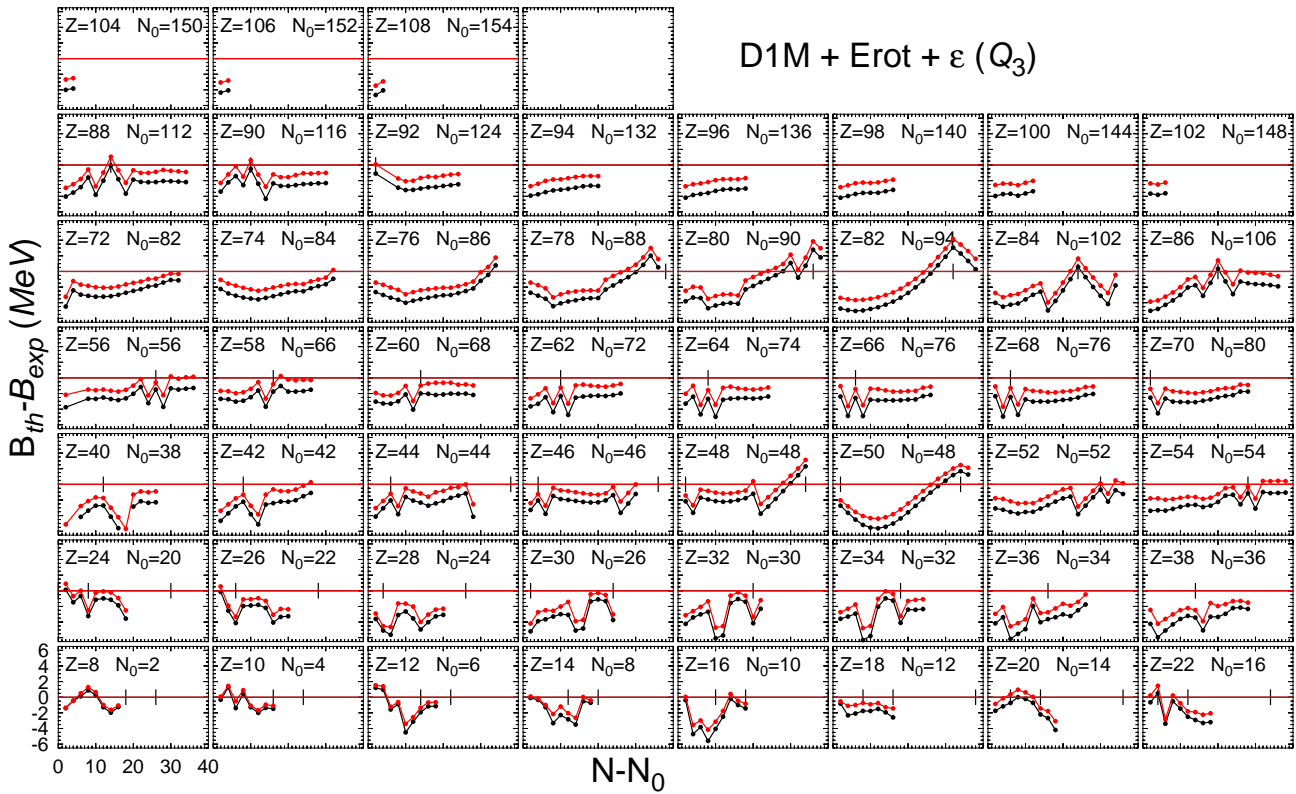}
\caption{
(Color online) Binding energy difference between the theoretical 
predictions and the experimental data (2012 compilation \cite{AW.12}) 
plotted as a function of neutron number differences N-N$_0$. The 
parameter N$_0$(Z) is a reference neutron number specific of each $Z$ 
value. In each panel the corresponding values of Z and N$_0$(Z) are 
given in the label. The horizontal line correspond to a perfect 
agreement theory-experiment. The vertical lines found along this reference 
horizontal line signal magic neutron numbers. The black curves 
correspond to the theoretical binding energy computed with D1M using 
the HFB energy plus the rotational energy correction. The red curves 
additionally include the octupole correlation energy of the GCM 
calculation. 
\label{fig-2}}
\end{figure*}
\begin{figure*}
\includegraphics[width=10.5cm,angle=270]{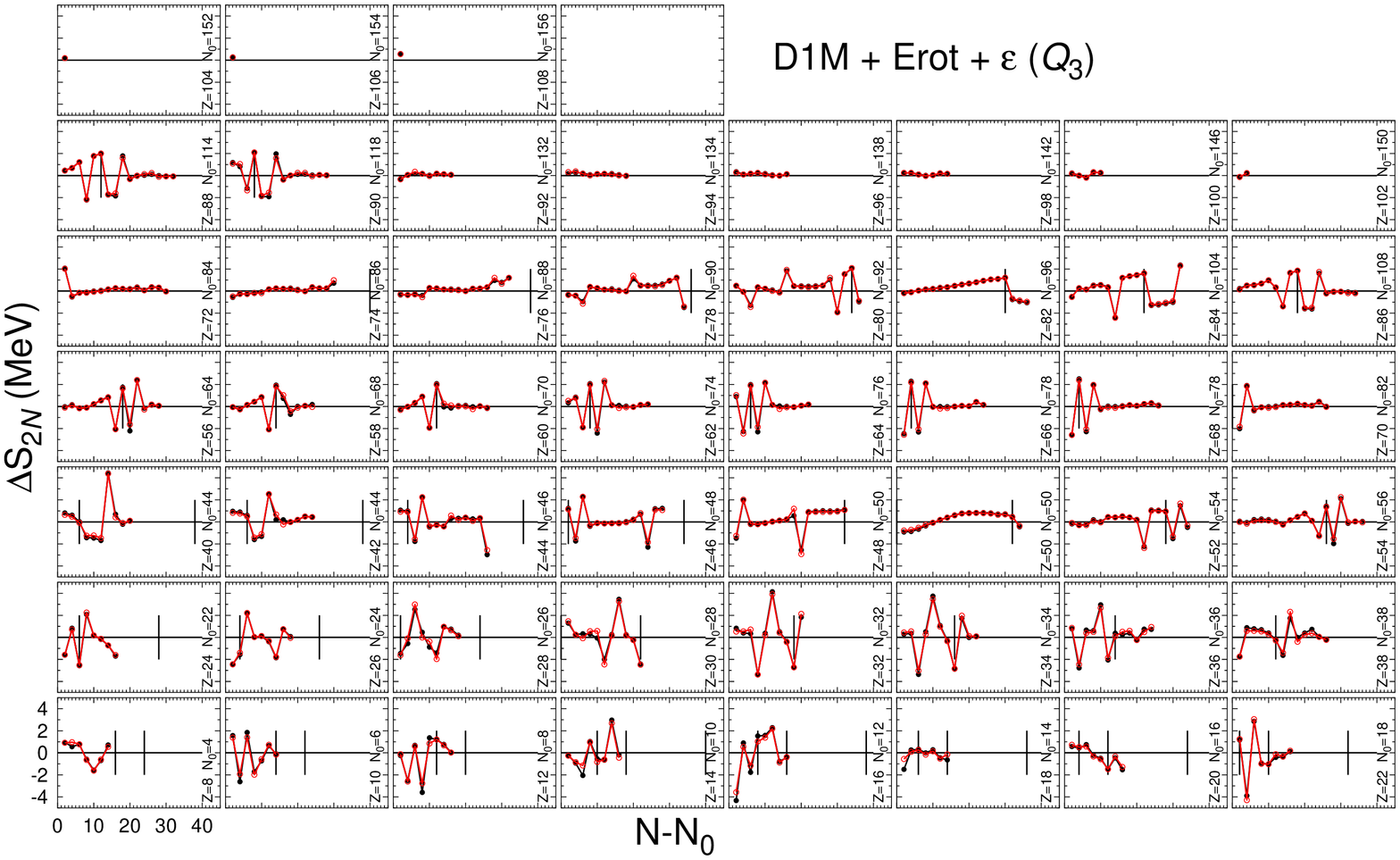}
\caption{
(Color online) The difference between the theoretical prediction for 
the two neutron separation energy and the experimental values $\Delta 
S_{2N}= S_{2N, th} - S_{2N, exp}$ is plotted for each Z value as a 
function of N-N$_0$ (see Fig \ref{fig-2}). Black (red) curves 
correspond to the results obtained with the HFB (GCM) theoretical 
approaches. 
\label{fig-3}}
\end{figure*}

First a HFB calculation preserving reflection symmetry is carried out 
for a set of 818 even even nuclei from oxygen to copernicium. The 
choice is made as to include the 620 even-even nuclei with known 
binding energies as given in the 2012 binding energy compilation 
\cite{AW.12} plus a few more. The binding energy obtained this way is 
used as the the reference energy in the definition of the different 
correlation energies. In Fig \ref{fig-1} the different correlation 
energies computed according to the approaches described above are 
plotted using a color scale. The three approaches are the reflection 
symmetry breaking HFB, the parity RVAP and finally the GCM with 
$Q_{30}$ as collective coordinate. In panel a)  the HFB results $\Delta 
E_\textrm{HFB}$ are depicted: just a few nuclei in the Ra, Ba and Zr 
regions show a non-zero correlation energy that never exceeds 1.2 MeV. 
Those are the regions where both proton and neutron numbers are close 
to the numbers favoring octupole correlations \cite{bu96}. Parity 
projection on top of both reflection symmetric as well as  octupole 
deformed HFB ground states produce a negligible energy gain as 
discussed in ref \cite{Rob.11b}. However, when parity RVAP is 
considered in addition to HFB, it leads to a non zero value of the octupole 
correlation energy in all the nuclei considered and as large as 1.5 
MeV, as shown in panel b). The parity RVAP produces the largest 
correlation energy increase in octupole soft nuclei whereas the increase is
essentially zero in those nuclei which are octupole deformed at the HFB 
level. The RVAP correlation energy shows some amount of shell effects as 
its value tend to be small close to magic proton and neutron numbers. 
Finally, the GCM correlation energy includes in its definition the 
symmetry breaking HFB correlation energy, the parity RVAP one as well 
as the correlation energy gained by the fluctuating octupole degree of 
freedom. This correlation energy is largest in those regions showing 
octupole deformation at the HFB level as observed in panel c) of  Fig 
\ref{fig-1}. The GCM correlation energy can be as large as 2.5 MeV. In 
the regions in between the correlation energy is not as large and is 
typically of the order of 1 MeV changing smoothly as a function of 
proton and neutron number and showing no indication of strong shell 
effects. At this point it can be concluded that ground state octupole 
correlation energies are not going to impact in a significant way the 
behavior of binding energies with proton or neutron number (only a 
roughly constant shift).

Similar results are obtained for the D1S and D1N \cite{D1N} 
parametrizations of the Gogny force. The corresponding plots can be 
found in the accompanying material. 

In order to understand the impact of the octupole correlation energy in 
the description of the binding energies we have plotted in Fig 
\ref{fig-2} the binding energy difference between two of the 
theoretical results and the experimental data (extracted from the 2012 
compilation \cite{AW.12}). These quantities are plotted  as a function 
of neutron number for each $Z$ value considered in the calculation 
(from oxygen to copernicium). In each panel two curves are shown: one 
corresponds to the HFB  with octupole deformation plus rotational 
correction results and the other to the results including in addition 
the octupole GCM correlation energy. In most of the cases, the 
inclusion of the GCM octupole correlation energy represents a mere 
displacement upwards of the curves and little improvement in the 
description of experimental data is observed. A comment is in order: in 
the calculations I do not have access to the zero point energy 
correction of the quadrupole degrees of freedom as computed in 
\cite{D1M} and therefore this quantity has not been added to the 
binding energy. It is typically of the order of 2-3 MeV and its neglect 
explains the overall shift observed in the D1M HFB predictions as 
compared to the experimental data. It is interesting to note that the 
largest discrepancies always take place around magic numbers and are 
specially remarkable for $Z=48-52$ and $Z=80-84$. In those cases the 
inclusion of the octupole correlation energy somehow reduces the 
discrepancy but is far from being enough to provide a good description 
of experimental data. When proton and neutron numbers are far from the 
magic numbers the behavior of the binding energy differences is quite 
smooth and almost independent of $N$ in most of the cases.

Another quantity connected to binding energies and of interest in 
astrophysics simulations is the two neutron separation energy $S_{2N}$. 
The reaction rates of nuclear reactions depends on the Q value  which 
is related to the $S_{2N}$. At magic neutron numbers this quantity 
shows a sudden drop with its magnitude known as the ``shell gap". The 
magnitudes of the shell gap for each proton number strongly influence 
the creation rate of heavy elements in nucleosynthesis environments. 
The shell gaps are not well reproduced by HFB mass models: before the 
drop of the $S_{2N}$ at magic neutron numbers the theoretical models 
predict  too high values for this quantity  but the prediction becomes too small 
after the drop. As a consequence the mean field values of the shell 
gaps are predicted to be too large. The expectation is that 
correlations beyond the mean field will  quench the "shell gap" (see 
\cite{Arc.12} for a discussion in the astrophysical context) but the 
amount of quenching associated to each of the effects considered on top 
of HFB is still under debate \cite{Tom.14}. To analyze the impact of 
octupole correlation energies on the shell gaps and the two-neutron 
separation energies the difference $\Delta S_{2N}=S_{2N, th} - S_{2N, 
exp}$ is plotted in Fig \ref{fig-3} in separated panels for each $Z$ 
value and as a function of $N$. As theoretical models I consider the 
HFB  and the octupole GCM. To help reading the results the zero is 
marked by a horizontal line and vertical lines are placed at magic 
neutron numbers. The too large mean field shell gaps are responsible for
the observed sudden change from positive to negative values when crossing magic neutron 
numbers. The conclusion extracted from the plot is that the 
extra correlation energies associated with the octupole GCM do not 
modify in a substantial way the HFB $\Delta S_{2N}$ values (the two 
curves are almost indistinguishable for most $Z$ values) and therefore 
octupole correlations do not help to improve the agreement with 
experimental data for the $S_{2N}$. As a consequence, octupole 
correlations do not quench the too large mean field "shell gaps". 

The present discussion has focused on the results for the D1M 
parametrization of the Gogny force but similar conclusions can be drawn 
for the D1S and D1N parametrizations. In the accompanying material 
plots like the ones of Fig \ref{fig-1} but for D1S and D1N are 
included.

I would like to point out that the irrelevant role played by octupole 
correlation energies in the description of nuclear binding energies and 
the associated reaction Q values by no means diminishes the importance 
of octupole correlations in the description of other aspects of nuclear 
structure like low lying negative parity  states and the associated 
enhanced E3 transition strengths (see \cite{bu96} for a thorough review 
including both phenomenology and theoretical methods, and \cite{Gaf.13} 
for a recent account), the physics of super-heavy nuclei 
\cite{Wan.12,Chen.08} or the description of alternating parity 
rotational bands \cite{Zhu.05,Gar.98} just to cite a few.


\section{Conclusions}

The ground state octupole correlation energy defined at three levels of 
approximation: the octupole deformed HFB, the parity RVAP and the 
octupole GCM are computed with the Gogny D1M effective interaction. 
Although non negligible and as large as 2.5 MeV, the octupole GCM 
correlation energy does not improve substantially the agreement between 
theory and experiment for binding energies. This is a consequence of 
the rather smooth behavior of the correlation energy as a function of 
$Z$ and $N$. The impact on the "shell gaps" is also imperceptible. I 
conclude that the ground state octupole correlation energy only 
contributes to minor improvements in the description of binding 
energies and plays a minor role in astrophysical simulations.

\begin{acknowledgments}
Work  supported in part by Spanish  MINECO grants Nos. FPA2012-34694 and FIS2012-34479
and by the Consolider-Ingenio 2010 program MULTIDARK CSD2009-00064.
\end{acknowledgments}

\end{document}